\documentclass[]{aa}
\usepackage{txfonts}
\usepackage{natbib}
\usepackage{graphicx}
\usepackage{longtable}
\bibpunct{(}{)}{;}{a}{}{,}
\begin{document}

\title{Spectroscopy of blue horizontal branch stars in NGC\,6656 (M22)
\thanks{Based on observations with the Blanco Telescope at Cerro
Tololo Inter-American Observatory (CTIO), Chile (program ID CN2010B-025)}
}

\author{
C. Salgado \inst{1}
\and
C. Moni Bidin \inst{1,2,3}
\and
S. Villanova \inst{1}
\and
D. Geisler \inst{1}
\and
M. Catelan \inst{3,4}
 }

\institute{
Departamento de Astronom\'ia, Universidad de Concepci\'on, Casilla 160-C, Concepci\'on, Chile
\and
Instituto de Astronom\'ia, Universidad Cat\'olica del Norte, Av. Angamos 0610, Antofagasta, Chile
\and
The Milky Way Milennium Nucleus, Av. Vicu\~na Mackenna 4860, 782-0436 Macul, Santiago, Chile
\and
Departamento de Astronom\'ia y Astrof\'isica, Pontificia Universidad Cat\'olica de Chile,
Av. Vicu\~na Mackenna~4860, 782-0436 Macul, Santiago, Chile
}

\date{Received / Accepted }


\abstract
{Recent investigations revealed very peculiar properties of blue horizontal branch (HB) stars in $\omega$\,Centauri, which show
anomalously low surface gravity and mass compared to other clusters and to theoretical models. $\omega$\,Centauri, however,
is a very unusual object, hosting a complex mix of multiple stellar populations with different metallicity and chemical abundances.}
{We measured the fundamental parameters (temperature, gravity, and surface helium abundance) of a sample of 71 blue HB stars in
M22, with the aim of clarifying if the peculiar results found in $\omega$\,Cen are unique to this cluster. M22 also hosts
multiple sub-populations of stars with a spread in metallicity, analogous to $\omega$\,Cen.}
{The stellar paremeters were measured on low-resolution spectra fitting the Balmer and helium lines with a grid of synthetic
spectra. From these parameters, the mass and reddening were estimated.}
{Our results on the gravities and masses agree well with theoretical expectations, matching the previous
measurements in three ``normal" clusters. The anomalies found in $\omega$\,Cen are not observed among our stars. A mild mass
underestimate is found for stars hotter than 14\,000~K, but an exact analogy with $\omega$\,Cen cannot be drawn.
We measured the reddening in the direction of M22 with two independent methods, finding $E(B-V)=0.35\pm0.02$~mag, with semi-amplitude
of the maximum variation $\Delta(E(B-V))=0.06$~mag, and an rms intrinsic dispersion of $\sigma(E(B-V))=0.03$~mag.}
{}


\authorrunning{Salgado et al.}
\mail{carosalgado@udec.cl}
\titlerunning{}
\maketitle


\section{Introduction}
\label{s_intro}

Horizontal branch (HB) stars in Galactic globular clusters (GCs) are old, low-initial mass (0.7--0.9 $M_{\sun}$) stars currently burning helium in their core \citep{Hoyle55,Faulkner66}. Owing to their complexity, some aspects of the formation and internal structure of HB stars have been a matter of debate for  decades \citep[see][for recent reviews]{Catelan09,Moni10}. For instance, the blue extension of the HB varies among clusters, a fact partly associated with the metallicity \citep{Sandage_wallerstein60} but not entirely explained by it \citep{Sandage67,VanDenBergh67}.
A number of second parameters in addition to metallicity have been suggested to provide an explanation for this behavior, but none has proven
completely adequate in describing the complex observational picture. Some examples are: stellar rotation \citep{Peterson83}, cluster concentration \citep{FusiPecci93}, presence of super-oxygen-poor stars \citep{Catelan95}, cluster mass \citep{RecioBlanco06}, environment of formation \citep{FraixBurnet09} and the cluster age \citep{Dotter10}. Many investigations showed that at least three parameters are required to describe the observations \citep[e.g.,][]{Buonanno97,Gratton10,Dotter10}.

The preponderance of recent evidence now  points towards the fact that two or more stellar generations can co-exist in the same cluster, with the younger stars being formed from material chemically enriched by the preceding generation \cite[e.g.,][]{Piotto09}. The most prominent case is undoubtedly $\omega$\,Centauri (NGC\,5139). This GC hosts a very complex mix of stellar sub-populations \citep{Bellini10}, and its double main sequence (MS) indicates the presence of a stellar generation extremely enriched in helium \citep[$Y$=0.38,][]{Norris04,Piotto05}. The helium abundance has long been proposed as a key parameter governing the cluster HB morphology \citep{DAntona02,Sweigart97}, because helium-rich HB stars are expected to be hotter than objects of canonical composition. This would in principle be consistent with the empirical correlation between HB morphology and presence of abundance anomalies that was first noted by \citet{Norris81} and \citet{Catelan95}.
Indeed, \citet{Lee05} claimed to have reproduced the complex HB behavour of $\omega$\,Cen assuming the presence of multiple stellar populations with different helium content. However, some problems still remains because, for example, both the HB gaps and hotter end of the HB in their model are $\sim$0.5~mag brighter than the observed data (see their Figure~3). The helium abundance has therefore been proposed as the third parameter governing the HB morphology, after metallicity and age (\citealt{Gratton12,Villanova12}; but see also \citealt{Catelan12}). Unfortunately, cool HB stars show no helium lines suitable for abundance measurements, while diffusion processes \citep{Michaud83,Michaud08,Quievy09} alter the surface chemical composition of HB stars hotter than the Grundahl jump \citep{Grundahl99}. As a consequence, reliable measurements of the primordial helium abundance of HB stars are suitable only in a very narrow range of temperature ($T_\mathrm{eff}$=9\,000--11\,000~K), where no helium-enriched stars are expected in metal-poor GCs \citep{Villanova09,Villanova12}. Nevertheless, an increased helium content can be indirectly deduced from other observable quantities, because He-enriched HB stars are brighter than their canonical counterparts, and occupy different loci in the temperature--gravity plane \citep[e.g.,][]{Moehler03,Catelan09,Catelan09b}.

\defcitealias{Moni11a}{MB11}
\citet[][hereafter MB11]{Moni11a} measured the temperature and gravity of a large sample of blue HB stars in $\omega$\,Cen, in search of indirect evidence of helium enrichment. Their measured gravities agree with the expectations for the He-enriched scenario, being systematically lower than the predictions of canonical models with solar helium abundance. However, this result cannot be straightforwardly interpreted as evidence of helium enrichment, because the resulting spectroscopic masses are unrealistically low. On the contrary, identical measurements for HB stars of three other comparison clusters revealed no such peculiarities. Hence, this behavior is so far unique to $\omega$\,Cen stars, and it has currently no explanation. A link between this result and the chemical peculiarities of this cluster can possibly be established by studying other GCs showing some similarity with $\omega$\,Cen.

\defcitealias{moni2007}{MB07}
\defcitealias{moni2009}{MB09}
\defcitealias{moni2012}{MB12}
In this paper, we investigate the properties of HB stars in M22 (NGC\,6656), measuring their fundamental parameters and comparing the results to those obtained in $\omega$\,Cen \citep[][hereafter MB12]{Moehler11,Moni11a,moni2012}, NGC6752 \citep[][hereafter MB07]{moni2007}, M80 (NGC\,6093) and NGC5986 \citep[][hereafter MB09]{moni2009}. M22 is a massive, metal-poor \citep[$\lbrack \mathrm{Fe/H}\rbrack\approx-1.8$,][]{marino_milone09} GC with an extended blue HB. It shows a multi-modal sub-giant branch \citep{marino2012}, and $\omega$\,Cen-like abundance variations but on a smaller scale \citep{dacosta2009}, as first suggested by \citet{hesser1977} and \citet{norris1983}. Revealing its chemical properties has proven difficult, because the differential reddening in the cluster area is substantial, and its color-magnitude diagram (CMD) is heavily contaminated by the bulge field population. A metallicity spread among its stars was thus debated for a long time, and it was eventually demonstrated by \citet{dacosta2009} and \citet{marino_milone09}, along with inhomogeneities of most of the chemical elements analyzed by the authors. Very recently, \citet{Marino13} studied a sample of seven HB stars in M22, supporting the idea that the HB-morphology is influenced by the presence of different stellar populations. Unfortunately, we do not have any target in common with their study for a direct comparison of the results.

\section{Observations and data reduction}
\label{s_obs}

\begin{figure}
\begin{center}
\includegraphics[width=7.8cm,angle=270]{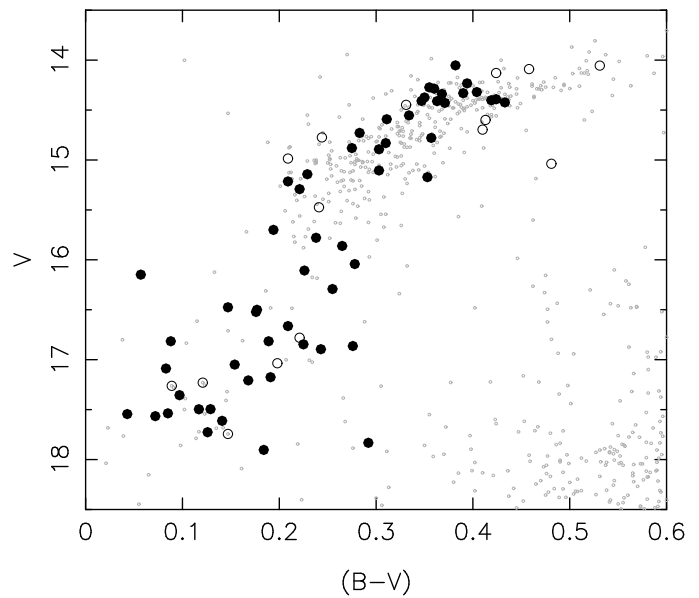}
\caption{Position of the target stars in the cluster CMD. The program stars are shown with large symbols, with empty dots used to indicate field objects, as discussed in the text. Grey, small dots are the cluster stars from chip~\#2 of \citet{monaco2004}.}
\label{f_cmd2}
\end{center}
\end{figure}

We selected 76 stars in M22 from the optical photometry of \citet{monaco2004}, distributed along the cluster HB from $T_\mathrm{eff}\sim$7\,000~K to $\sim$30\,000~K. The position of the targets on the HB is presented in Fig.~\ref{f_cmd2}, where small dots are used to show all the sources from chip~\#2 of Monaco et al's photometry, that includes the cluster central region but not all our targets. In Table~\ref{Tabla1} we give the IDs, magnitudes $V$, and colors ($B-V$) from the \citet{monaco2004} database. The IDs are given in the form $x$-$yyyy$, where $x$ is the chip number in Monaco et al.'s photometry, and $yyyy$ is the star number in the corresponding catalog.

The data were collected at CTIO, during one night of observations (August 27, 2010) with the Hydra optical spectrograph mounted on the Blanco 4.0m telescope. We used the grating KPGL1, coupled with the BG39 filter and the 200$\mu$ slit plate, to cover a wavelength range of 3700--4900~\AA\ with a spectral resolution of 2.7~\AA . All the program stars were targeted with one fiber configuration only, and 10 fibers were allocated to the sky background. Six 1200s exposures were collected, and reduced independently. The spectra of two targets were of too low quality to be used, and they were discarded.
The frames were de-biased and flat-fielded with standard IRAF\footnote{\small{IRAF is distributed by the National Optical Astronomy Observatories, which are operated by the Association of Universities for Research in Astronomy, Inc., under cooperative agreement with the National Science Foundation.}} routines, using the calibration frames collected during daytime operations. The spectra were then extracted and wavelength-calibrated with the {\it dohydra} IRAF task, by means of the arc lamp image collected in the middle of the six exposures. For each frame, the sky fibers were extracted, averaged, and subtracted to the science spectra.
The spectrum of the standard star EG274 \citep{hamuy1992}, acquired during the same night, was used to derive the instrumental response curve. Unfortunately, this procedure failed to return the correct stellar continuum of the program stars. Since the standard star is particularly bright and its flux close to the CCD saturation, we suspect that this was due to a loss of linearity of the detector response. We therefore normalized the science spectra fitting a low-order polynomial function to the stellar continuum.

\begin{figure}
\begin{center}
\includegraphics[width=6.9cm,angle=270]{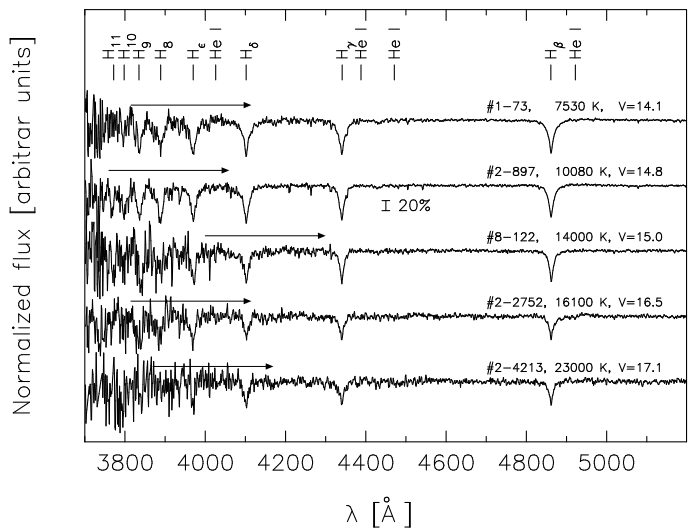}
\caption{Examples of the normalized spectra of target stars at different temperatures. The spectra were vertically shifted to avoid overlap. The Balmer and helium lines and the spectral range of each star (black arrows) used in our fitting routine are indicated.}
\label{f_spectra}
\end{center}
\end{figure}

We measured the heliocentric radial velocity (RV) of the target stars through IRAF's task {\it fxcor}, which employes the cross-correlation method described in \citet{tonry&davis1979}, adopting as template a synthetic spectrum of a typical HB star, with parameters similar to the program stars. The exact choice of the template is unimportant in terms of the RV measurements, because a mismatch between the template and object parameters increases the formal uncertainties but does not introduce systematic errors \citep{morse1991,moni2011b}.
The RV of the targets was measured in each of the six spectra independently, to check for consistency. In three cases, we found large discrepancies between the measurements ($>100$~km~s$^{-1}$), incompatible with the observational errors. All the data were collected within two hours only, and binaries are rare among HB stars of GCs \citep{moni2006,moni2008}, hence these RV variations are unlikely to be real. We interpreted them as a sign of problems with the spectra, and these three stars were excluded from the analysis. The six spectra of each of the remaining 71 targets were summed. Examples of the final co-added spectra are presented in Fig.~\ref{f_spectra}. The typical signal-to-noise ratio varied from $\sim$120 for the brightest stars to $\sim$20 for the hottest targets. The RV of each object was eventually determined on the final co-added spectrum, whose higher quality permitted a better accuracy. The typical error deriving from the cross-correlation procedure was $\approx$18~km~s$^{-1}$. The results are provided in the last column of Table~\ref{Tabla1}.

\begin{figure}
\begin{center}
\includegraphics[width=6.2cm,angle=270]{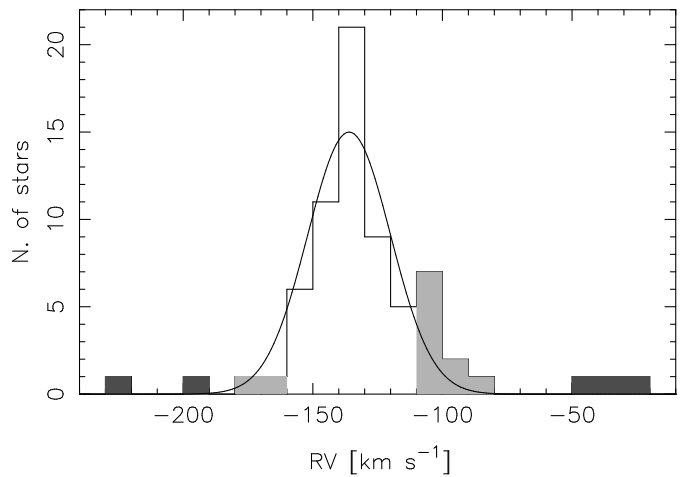}
\caption{Distribution of the measured RVs. The black area denotes the targets flagged as non-cluster members due to discrepant RV, while the grey area indicates those whose membership is uncertain. The solid curve shows a Gaussian distribution with $\sigma\approx$18~km~s$^{-1}$ centered on RV=$-136$~km~s$^{-1}$.}
\label{f_histo}
\end{center}
\end{figure}

The RV distribution of the target stars is shown in Fig.~\ref{f_histo}. A peak is found at RV$\approx -136$~km~s$^{-1}$, which is $\sim 10$~km~s$^{-1}$ higher than the nominal cluster RV \citep{marino_milone09,Peterson1994}. We deduce that our measurements could be affected by a systematic offset, which is still negligible compared to the resolution of our spectra (FWHM=2.7~\AA $\approx$170~km~s$^{-1}$). The distribution is not symmetric, because an overdensity of stars, probably a second peak, is found at RV$\approx-100$~km~s$^{-1}$. A few objects with RV very different from the mean value ($\Delta(RV)>$50~km~s$^{-1}$) are also found on both sides of the main peak. While measurement errors could cause these very discrepant values, the asymmetry of the distribution is a clear sign of field contamination. On the other hand, given our large measurement errors and the kinematical complexity of the bulge field in the direction of M22, the field and cluster populations cannot be easily disentangled on the basis of RV only: as shown in Fig.~\ref{f_histo}, some cluster stars are expected even at RV$\sim-100$~km~s$^{-1}$, where the field contamination apparently dominates. However, field contaminants are expected to show some peculiarity in addition to a deviating RV, because MS background stars should have higher gravity and mass and no helium depletion, while the resulting mass of field subdwarfs could be biased if their distance and/or reddening differ from the assumed cluster values. For these reasons, we cleaned the sample from field contamination by means of the following conservative criteria: basing only on RVs, we identify as field objects the stars whose RV is outside the interval $-190<$RV$<-80$~km~s$^{-1}$, because Fig.~\ref{f_histo} shows that no cluster member is expected outside this range. These are indicated with a black dot after the RV in Table~\ref{Tabla1}, and a black area in the histogram of Fig.~\ref{f_histo}. The stars with $-190<$RV$<-160$~km~s$^{-1}$ and $-110<$RV$<-80$~km~s$^{-1}$ are flagged as ``uncertain cluster membership", indicated with a grey area in Fig.~\ref{f_histo}, and with an empty circle in Table~\ref{Tabla1}. The same symbol is used to mark the stars with deviating gravity, mass, or helium abundance, as discussed in Sect.~\ref{ss_temp_grav}, \ref{ss_masses}, and \ref{ss_abundance}, respectively. Any star whose membership is defined uncertain twice or more is then identified as a field object. All field contaminants are marked with ``F" in the last column of Table~\ref{Tabla1}, indicated with empty symbols in all the figures, and excluded from the analysis.


\section{Measurements}
\label{s_meas}

For the purpose of measuring the effective temperature, surface gravities, and helium abundance of the target stars, we fitted the hydogen and helium lines with synthetic spectra. Stellar model atmospheres computed with ATLAS9 \citep{kurucz1993} were used as input of Lemke's version\footnote{\small{http://a400.sternwarte.uni-erlangen.de/\~{}ai26/linfit/linfor.html}} of the Linfor program (developed originally by Holweger, Steffen and Steenbock at Kiel University), to create a grid of synthetic spectra covering the range 7\,000$\leq T_\mathrm{eff} (K)\leq$35\,000, $-$3.0$\leq \log{N(He)/N(H)} \leq$ $-$1.0, and gravity increasing with temperature from 2.5$\leq \log{g} \leq$ 5.0 at the cooler end to 5.0$\leq \log{g} \leq$6.0 for the hotter stars. The calculation included the Balmer lines H$_\alpha$ to H$_{22}$, four $\ion{He}{I}$ lines (4026~\AA, 4388~\AA, 4471~\AA, 4922~\AA), and two $\ion{He}{II}$ lines (4542~\AA, 4686~\AA). The synthetic spectra  were convoluted with a Gaussian (FWHM=2.7\AA), to match the instrumental resolution of the data.

The targets were classified in two groups, according to their temperature, to determine the set of models to be used in the fitting routine. The stars with 
$T_\mathrm{eff} \geq$ 12\,500\,~K, as deduced by their position in the CMD, were fitted with super-solar metallicity models ([M/H]=$+$0.5) and variable surface helium abundance. On the other hand, the coldest  targets in the sample ($T_\mathrm{eff} \leq$11\,000\,~K) were treated with models with metallicity [M/H]=$-$1.5 whose helium abundance was kept fixed at the solar value. The spectra of stars across the Grundahl jump \citep{Grundahl99}, with a color temperature in the range 11\,000$-$12\,500 K, were analyzed in search of evidence of active atmospheric diffusion. The stars presenting significant $\ion{Fe}{II}$ lines in the range 4450--4600~$\AA$ (\citealt{moehler1999}; \citetalias{moni2007}), or clearly lying blue-ward of the jump in the CMD, were treated as the hot group and fitted with super-solar metallicity models, while the others were fitted with metal-poor models as the cool targets.

The best fit and the derivation of the stellar parameters were established through the routines developed by \citet{bergeron1992} and \citet{saffer1994}, as modified by \citet{napiwotzki1999}. The code normalizes the object and the synthetic spectra simultanously using the same points for the continuum definition, and it makes use of a $\chi^{2}$ test, estimating the required $\sigma$ from the noise in the continuum spectrum. The lines of the Balmer series from H$_{\beta}$ to H$_{8}$ (excluding H$_{\epsilon}$ to avoid the blended $\ion{Ca}{II}$~H line), were always included in the fitting routine, along with the four $\ion{He}{I}$ lines for stars whose helium abundance was a free parameter. These lines were too faint in the spectra of cooler stars to be analyzed at our spectral quality and resolution. We however verified that their inclusion in the routine had no effect on the results, and that these lines were consistent with the prediction of the resulting best-fit synthetic spectrum with solar helium abundance.
The hydrogen lines H$_{9}$, H$_{10}$, and H$_{11}$, falling in the very noisy bluer range, were employed in the fit when clearly visible, usually for the cooler, brighter stars. The two $\ion{He}{II}$ lines were also fitted in the spectra of stars with $T_\mathrm{eff}>$30\,000~K, while they were not detected for cooler targets.

The results are shown in  Table~\ref{Tabla1}. 
The fitting routine estimates the errors based only on the statistical noise of the spectrum, and other sources of uncertainty (e.g., normalization, sky-subtraction, flat-fielding) are neglected. The code thus likely underestimates the true uncertainties by a factor of 2--4 (R. Napiwotzki 2005, priv. comm.). We  therefore multiplied the formal errors by three, to allow for a more realistic approximation of the true uncertainties. The errors thus modified are quoted in Table~\ref{Tabla1}.

The stellar masses were estimated from the parameters, making use of the equation
\begin{equation}
\log{\frac{M}{M_{\sun}}}=\log{\frac{g}{g_{\sun}}}-4\cdot \log{\frac{T_\mathrm{eff}}{T_\mathrm{eff,\sun}}}+\log{\frac{L}{L_{\sun}}},
\label{eq_mass1}
\end{equation}
where
\begin{equation}
\log{\frac{L}{L_{\sun}}}=-0.4\cdot(V - (m-M)_V + BC_V - M_\mathrm{bol,\sun}).
\label{eq_mass2}
\end{equation}
The bolometric correction (BC$_V$) was derived from the effective temperature through the empirical calibration of \citet{flower1996}. We assumed $T_{\sun}$=5777 K, $\log{g_{\sun}}$=4.44, $(m-M)_V$=13.74$\pm$0.20 \citep{monaco2004}, and $M_\mathrm{bol,\sun}$=4.75, because $M_{V,\sun}$=4.83 \citep{Binney98} and the \citet{flower1996} BC$_V$--$T_\mathrm{eff}$ relation returns BC$_{\sun}=-0.08$. The errors in masses were derived using standard propagation theory.


\section{Results}
\label{s_res}

\subsection{Temperatures and gravities}
\label{ss_temp_grav}

\begin{figure}
\begin{center}
\includegraphics[width=14cm,angle=270]{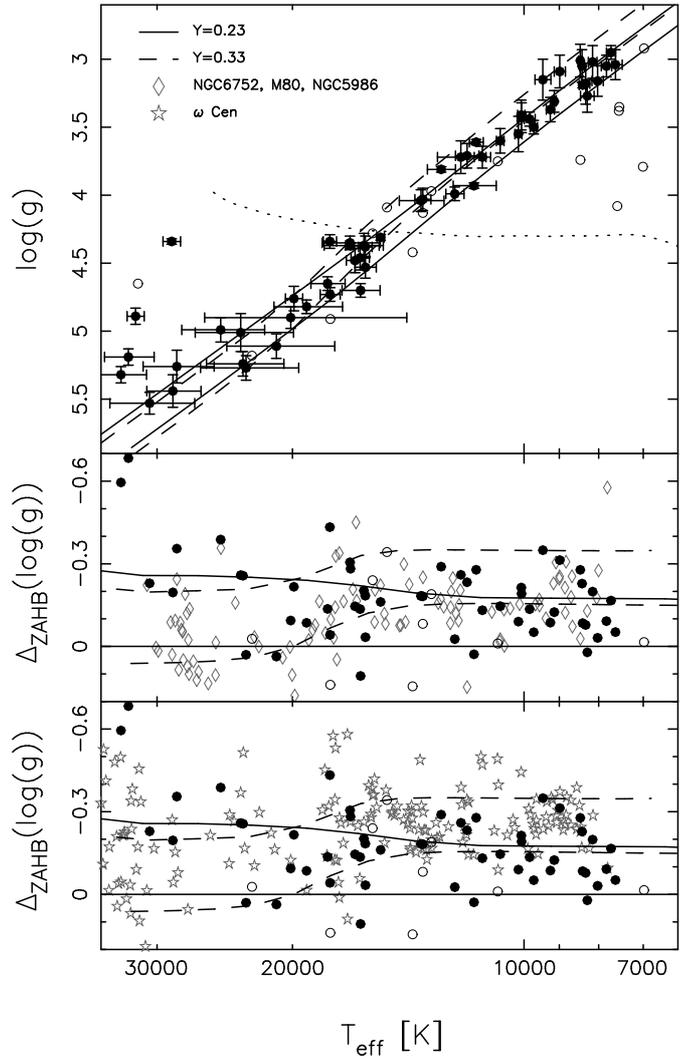}
\caption{Upper panel: position of the targets in the temperature-gravity plane. Empty and full circles are as in Fig.~\ref{f_cmd2}. The canonical zero-age and terminal-age HB for canonical stars are shown with full lines, and for analogous helium-enriched (Y=0.33) models are indicated with dashed lines. The zero-age main sequence of solar chemical composition of \citet{Salasnich00} is shown as dotted line. Middle and lower panel: comparison of stars in M22 (same symbols as the upper panel) and members of NGC\,6752, M80, and NGC\,5986 (open, grey diamonds), and $\omega$\,Cen (open, grey stars). The vertical coordinate is the difference between the stellar gravity and the corresponding value of the canonical ZAHB at the same temperature. The full and dashed curves are the same as in the upper panel.}
\label{logt_logg}
\end{center}
\end{figure}

The location of the targets in the temperature-gravity plane is shown in the upper panel of Fig.~\ref{logt_logg}. The general trend of the stars with $T_\mathrm{eff}<25\,000$~K ($\log{(T_\mathrm{eff})}<4.4$) agrees well with the expectations of the canonical models. The majority of these stars are found between the canonical zero-age and terminal-age HB (ZAHB and TAHB, respectively), where helium-burning stars spend 99\% of their lifetime. A mild underestimate of gravities can be seen, as the stars scatter mainly toward lower gravities. This behavior, however, was already noted by \citetalias{Moni11a} in the non-peculiar clusters NGC\,6752, M80, and NGC\,5986. As an exception, five objects at the coolest end of the temperature distribution deviate from the general trend due to an excessive gravity. These are likely field MS stars, although their $\log{(g)}$ is lower than the expectation of the zero-age MS model shown in Fig.~\ref{logt_logg}. This can be in part expected, because massive stars evolve toward lower gravities during their MS lifetime. Disregarding gravities, and assuming for each of these targets the theoretical magnitude and color of a star at the same temperature from the zero-age MS of solar chemical composition of \citet{Salasnich00}, we obtain the consistent picture where they are all foreground objects between 1.2 and 2.3~kpc from the Sun, with reddening increasing with distance, from $E(B-V)$=0.09 to 0.25~mag \citep[$d$=3.2~kpc and $E(B-V)$=0.38~mag for M22][]{monaco2004}. They are consequently flagged as ``uncertain cluster membership" objects in Table~\ref{Tabla1} and, according to the criteria given in Sect.~\ref{s_obs}, they are all eventually classified as field objects.

The gravity of the eight targets with $T_\mathrm{eff}>25\,000$~K is systematically lower than the model expectations. The quality of the spectra of these faint stars is low (S/N$\approx$20), and this could have induced an incorrect placement of the continuum in the normalization procedure, biasing the measurements. However, the low gravity of these objects could also indicate that they are post-helium burning stars evolving off the HB \citep[see, e.g.,][]{Moehler11}.

In the middle and lower panel of Fig.~\ref{logt_logg}, we compare the results for the program stars with previous measurements in $\omega$\,Cen, NGC\,6752, M80, and NGC\,5986. We adopt as vertical coordinate the difference between the measured gravity and the value of the canonical ZAHB at the corresponding temperature, $\Delta_{\rm ZAHB}=\log{(g)}-\log{(g)}_{\rm ZAHB}$. The plot is thus analogous to the temperature-gravity space of the upper panel, but the horizontal axis coincides with the canonical ZAHB. The plot reveals an excellent match between our results in M22 and in the three  ``normal" comparison clusters, at variance with $\omega$\,Cen. The mean offset of HB stars with $T_\mathrm{eff}<25\,000$~K in M22 is $\overline{\Delta_\mathrm{ZAHB}\log{(g)}}=-0.15\pm0.02$~dex, constant in the whole temperature range. By comparison, we find $\overline{\Delta_\mathrm{ZAHB}\log{(g)}}=-0.14\pm0.02$~dex in the three comparison  clusters, slightly decreasing from $-0.16\pm0.02$~dex for $T_\mathrm{eff}\leq14\,000$~K to $-0.10\pm0.03$~dex in the range 14\,000--25\,000~K. On the contrary, $\omega$\,Cen HB stars show an offset of $-0.27\pm0.02$~dex, constant with temperature, with respect to the canonical ZAHB. Hence, HB stars in M22 do not show the low gravities observed in $\omega$\,Cen, which remains a peculiarity of this extraordinary object.

\subsection{Masses}
\label{ss_masses}

\begin{figure}
\begin{center}
\includegraphics[width=8.3cm,angle=270]{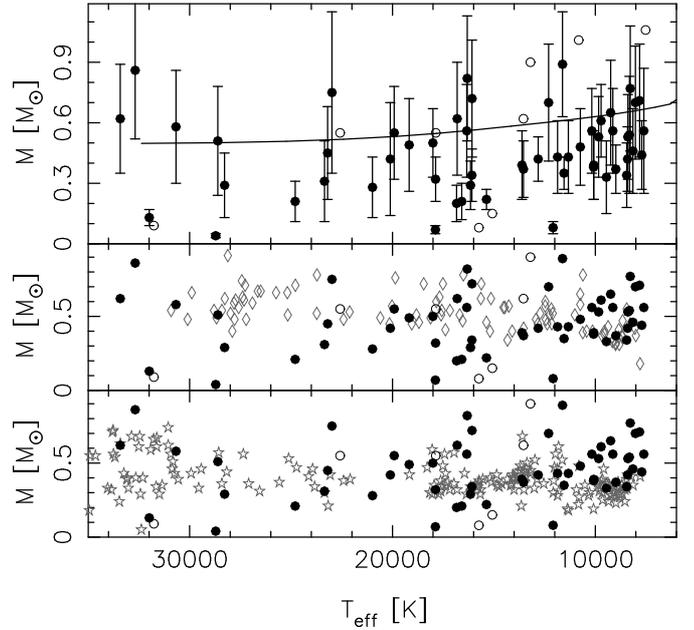}
\caption{Spectroscopic masses of the program stars (full and empty circles, as in Fig.~\ref{f_cmd2}) as a function of temperature, compared to the expectation of the canonical ZAHB model (line in the upper panel), and to the results for NGC\,6752, M80, and NGC\,5986 (grey diamonds in the middle panel), and $\omega$\,Cen (grey stars in the lower panel).}
\label{logt_masa}
\end{center}
\end{figure}

The spectroscopic mass of our targets is plotted in Fig.~\ref{logt_masa} as a function of temperature, and compared to the ZAHB expectations. The masses of HB stars in GCs are well established by stellar evolution theories. In particular, they cannot be lower than the He-core mass at the helium flash ($\sim0.47$~M$_{\sun}$), nor higher than the turnoff mass ($\sim$0.8~M$_{\sun}$), and any result beyond these constraints evidences problematic measurements. However, it must be noted that the calculation of the mass requires the knowledge of the distance and reddening, for which we assumed the cluster values. As a consequence, the result can be highly biased for non-cluster members. We therefore identified as field contaminants (marked with a black dot after the mass column of Table~\ref{Tabla1}) the eight stars with M$>1$~M$_{\sun}$. Five of these stars also show peculiar RV or gravity. We also note that no star among the 115 cluster members studied by \citepalias{Moni11a} show a spectroscopic mass lower than 0.18~M$_{\sun}$, despite the general large underestimate. Hence, we will conservatively flag the five objects with M$<0.1$~M$_{\sun}$ as ``uncertain cluster membership" (marked with a circle in Table~\ref{Tabla1}).

Figure~\ref{logt_masa} reveals that the masses of HB stars in M22 are slightly underestimated compared to theoretical expectations, but they show a good agreement with the previous estimates in NGC\,6752, M80, and NGC\,5986. In fact, the average mass in M22 and in the three comparison clusters is 0.46$\pm$0.03 and 0.49$\pm$0.01~M$_{\sun}$ respectively, in both cases lower than the 0.57~M$_{\sun}$ expected on the ZAHB. We note that a slight difference could be expected, because evolution at the blue HB always proceeds redwards, meaning that at any temperature there will certainly be stars with lower masses that originated from a hotter ZAHB position. The effect is likely small though. The trend with temperature is not the same in M22 and in the other GCs. The masses of the stars cooler than 14\,000~K are on average 0.07$\pm$0.04~M$_{\sun}$ higher in M22 than in the other clusters. The theoretical expectation decreases with temperature, and the mean offset of M22 stars with respect to the model remains constant over the full range. On the contrary, the mean mass in the three reference clusters increases, and $\overline{\mathrm{M}}=0.56\pm$0.02~M$_{\sun}$ for stars hotter than 14\,000~K, where the model expectation is 0.54~M$_{\sun}$, and the average mass of M22 stars is $\overline{\mathrm{M}}=0.40\pm0.04$~M$_{\sun}$. By comparison, the average value found in $\omega$\,Cen is $0.36\pm$0.01~M$_{\sun}$, constant over the whole temperature range. The uncertainty in the distance modulus of M22 can partially explain the mass underestimate, because masses higher by 0.08~M$_{\sun}$ are obtained if $(m-M)_V$ is increased by 1$\sigma$=0.2~mag. However, both the mean mass underestimate in M22 and the different trend with temperature in the three comparison clusters are fully explained by the small offset in gravity $\overline{\Delta_\mathrm{ZAHB}\log{(g)}}$ discussed in Sect.~\ref{ss_temp_grav}, and by its different dependence on temperature.

In conclusion, the mass underestimate found in $\omega$\,Cen is not observed in M22, at least among stars cooler than 14\,000~K, where the results are similar to those obtained in NGC\,6752, M80, and NGC\,5986. A mild mass underestimate could still be present among stars hotter than this temperature, where the average mass of M22 stars is intermediate between $\omega$\,Cen and the other clusters. However, this small offset is fully accounted for by the lower gravities, at variance with $\omega$\,Cen stars \citepalias{Moni11a}. Moreover, while the average value in this temperature range is affected by a group of stars with low mass at $T_\mathrm{eff}\approx$16\,000~K, the general trend observed in Fig.~\ref{logt_masa} does not suggest any analogy with $\omega$\,Cen.

The average mass of the eight targets hotter than 25\,000~K is consistent with that of the cooler stars, and an evident underestimate is not observed. This is at variance with what we would have expected if their low gravity (Sect.~\ref{ss_temp_grav}) were a product of an incorrect normalization of the spectra. \citetalias{moni2007} discovered a group of hot stars in NGC\,6752 with anomalously high spectroscopic mass, which possibly occupy redder and fainter loci in the CMD. They were later identified even in M80 \citepalias{moni2009} and $\omega$\,Cen \citepalias{moni2012}. So far they have been found only at $T_\mathrm{eff}\geq$25\,000~K. We detect one hot star with spectroscopic mass higher than the theoretical expectations by more than 1$\sigma$, but this is neither redder nor fainter than the bulk of the HB population ($V$=16.15, ($B-V$)=0.06, see Fig.~\ref{f_cmd2}). Unfortunately, our sample comprises only eight targets in this temperature range, which is insufficient to draw firm conclusions about the presence of these high-mass objects in M22. Their frequency is unknown, with estimates varying from 20\% in $\omega$\,Cen to 40\% in NGC\,6752, hence only 1--3 such objects would have been expected in our sample, with a non-negligible probability of a null detection.

\subsection{Helium abundance}
\label{ss_abundance}

\begin{figure}
\begin{center}
\includegraphics[width=6.cm,angle=270]{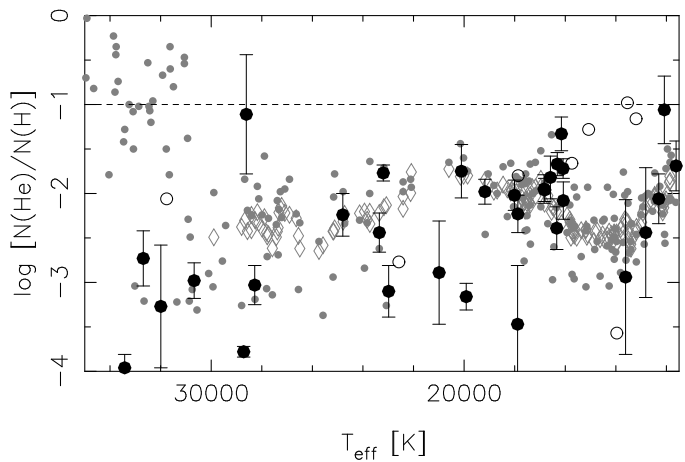}
\caption{Surface helium abundance of targets stars (empty and full symbols with errorbars, as in Fig.~\ref{f_cmd2}), as a function of their temperature. Small, grey dots indicate the results in NGC\,6752, M80, NGC\,5986, and $\omega$\,Cen. The grey diamonds show the trend of helium abundance with temperature derived by \citetalias{moni2012}. The dotted line indicates the solar abundance.}
\label{logt_abun}
\end{center}
\end{figure}

The surface helium abundance derived for our target stars is plotted in Fig.~\ref{logt_abun} as a function of the effective temperature, where we compare the results with the measurements in the other clusters. As discussed by \citetalias{moni2012}, the diffusion processes completely alter the surface abundances of blue HB stars as compared to their pre-HB values, and the same behavior is observed in all the clusters. Our measurements reveal that M22 is no exception, and our stars closely follow the trend with temperature described in \citetalias{moni2012}. The results for the targets hotter than 20\,000~K are consistent with the presence of two groups of stars, with mean abundance $\log{(N(\mathrm{He})/N(\mathrm{H}))}\approx-2$ and $-3$~dex respectively, as found in the other clusters \citepalias{moni2012}, although the sample comprises too few objects to draw firm conclusions.

The helium abundance of three targets is close to the solar value, and higher by more than 2$\sigma$ from the mean abundance of the other clusters, suggesting that they could be field contaminants. In fact, MS stars are usually not affected by atmospheric diffusion. Unfortunately, the temperature and gravity of these targets do not help clarify their nature, because they are found in a region of the $T_\mathrm{eff}$--$\log{g}$ plane where the ZAMS crosses the HB tracks (13\,000--15\,000~K, $\approx$4.1~dex, see Fig.~\ref{logt_logg}). However, they all show discrepant RV, and the spectroscopic mass in two cases is barely compatible with the exclusion criteria defined in Sect.~\ref{ss_masses}. These objects were flagged as ``uncertain cluster membership" and, following the criteria of Sect.~\ref{s_obs}, consequently identified as field stars due to their RV.

Our sample reaches the canonical end of the HB at 32\,000~K. The five hottest targets can be, within the errors, hotter than this limit, and they are therefore candidates for blue hook (BH) stars \citep{Whitney98,DCruz00,Brown01}, whose formation cannot be explained by canonical stellar evolution theories. However, \citet{Moehler11} showed that genuine BH objects have solar or super-solar surface helium abundance, and $\omega$\,Cen BH stars are visible at the hottest end of Fig.~\ref{logt_abun}, at $\log{(N(\mathrm{He})/N(\mathrm{H}))}\approx-1$. On the contrary, all our hottest targets are strongly depleted in helium ($\log{(N(\mathrm{He})/N(\mathrm{H}))}<-2$), and they are therefore more likely post-helium burning stars evolving off the HB toward the white dwarf cooling sequence.

\subsection{Reddening}
\label{ss_reddening}

\begin{figure}
\begin{center}
\includegraphics[width=4.8cm,angle=270]{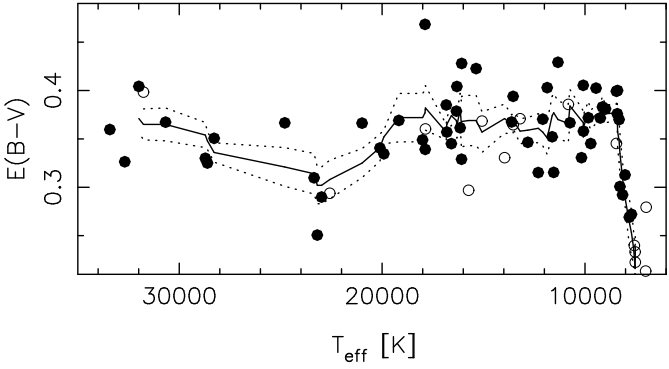}
\caption{Reddening of the program stars as a function of their temperature. Empty and full circles are used as in Fig.~\ref{f_cmd2}. The full line indicates the trend of the mean value, calculated as described in the text, with its 1$\sigma$ range (dotted line).}
\label{logt_reddening}
\end{center}
\end{figure}

The reddening of each target was estimated by comparing the observed color $(B-V)$ with the theoretical color of a star at the same temperature and gravity. This was calculated interpolating the grid of \citet{kurucz1993}, adopting the same metallicity as in the spectra fitting process. The results are plotted in Fig.~\ref{logt_reddening}, where the full and dotted lines show the trend of the mean reddening and its 1$\sigma$ range, respectively. This was calculated substituting for each star the average of the five adjacent targets in order of temperature. The mean reddening of the stars hotter than 8\,000~K is $\overline{E(B-V)}=0.36\pm0.01$~mag, with no trend with temperature. This result agrees well with previous measurements in the literature \citep{Zinn85,Webbink85,Reed88}, whose mean value reported by \citet[][2010 web version]{harris1996} is $E(B-V)$=0.34~mag. In particular, our result is consistent with $E(B-V)=0.38\pm0.02$~mag found by \citet{monaco2004}, whose photometric data were employed in this work. This good agreement with previous estimates suggests that the temperature scale derived in our study is correct within 10\%, because a systematic error of this magnitude would change the derived reddening by $\pm0.02$~mag.

The stars span an interval with semiamplitude $\Delta(E(B-V))=\pm0.06$~mag, and the standard deviation of our measurements is $\sigma(E(B-V))=0.033$~mag. Quadratically subtracting the typical photometric error for our stars, and the uncertainty propagated by the errors on the parameters ($\sim$0.01~mag in both cases), we obtain an intrinsic dispersion of $\sigma(E(B-V))=0.03$~mag. These values are remarkably similar to those quoted by \citet{monaco2004}: $\sigma(E(B-V))=0.02$ and $\Delta(E(B-V))=0.06$~mag. Our results agree well with \citet{AnthonyTwarog95} and \citet{Richter99} also, who quoted $\Delta(E(B-V))=0.07$ and 0.08~mag, respectively, and with $\sigma(E(V-I))=0.05$~mag by \citet{Piotto99HST}, because our measurement translates into $\sigma(E(V-I))=1.34\cdot\sigma(E(B-V))=0.04$~mag by means of the transformations of \citet{Cardelli89}.

A reddening underestimate appears evident for stars cooler than $\sim$8\,000~K, where $\overline{E(B-V)}=0.26\pm0.02$~mag. Most of the targets in this temperature range are flagged as field objects, and the reddening of foreground stars could indeed be intrinsically lower. The possible gravity underestimate for MS stars discussed in Sect.~\ref{ss_temp_grav} should play a minor role, because gravity has a negligible impact on the theoretical color. Nevertheless, ascribing the reddening underestimate at the coolest edge of our sample only to field contamination is unsatisfactory, because the same trend is common to all the stars, irrespective of their cluster membership status. Unfortunately, we cannot compare our results with those in the other clusters, because previous investigations did not target such cool stars, although a mild tendency toward a lower reddening is found among the coolest stars of \citetalias{moni2007}.

It is possible that the metal lines, unaccounted for in the model spectra, could be strong enough to alter the line profiles in the spectra of these cool stars, thus biasing the results. We checked this issue repeating the measurements on a high-resolution synthetic spectra with $T_\mathrm{eff}$=7\,000--7\,500~K and $\log{g}$=3~dex extracted from the library of \citet{Coelho05} and convolved with a Gaussian profile to match the resolution of our data. We indeed found a temperature underestimate of $\approx$10\% on these spectra. However, the temperature should be increased by 2\,000--2\,500~K to increase the reddening by $\sim$0.10~mag, and the spectroscopic mass would consequently decrease by $\sim$0.3~M$_{\sun}$ on average. This would lead to unphysical results, far from both the model expectations and the trend of the cluster stars (see Fig.~\ref{logt_logg} and~\ref{logt_masa}). Hence, the reddening underestimate cannot be easily ascribed to a bias in the derived temperatures. The problem could also reside in the theoretical colors. We note that empirical calibrations of temperature-color relations show that synthetic colors from model atmospheres calculated with ATLAS9 have difficulty in properly reproducing the observations of cool stars \citep{Ramirez05,Sekiguchi00}, a problem also pointed out in the theoretical investigation by \citet{Castelli97}. Our results do not, by themselves, demonstrate the inadequacy of these models for HB stars at $T_\mathrm{eff}\approx7\,000~K$, but they hint that the observed reddening underestimate could be part of the same phenomenon evidenced by other authors.

\begin{figure}
\begin{center}
\includegraphics[width=9.3cm]{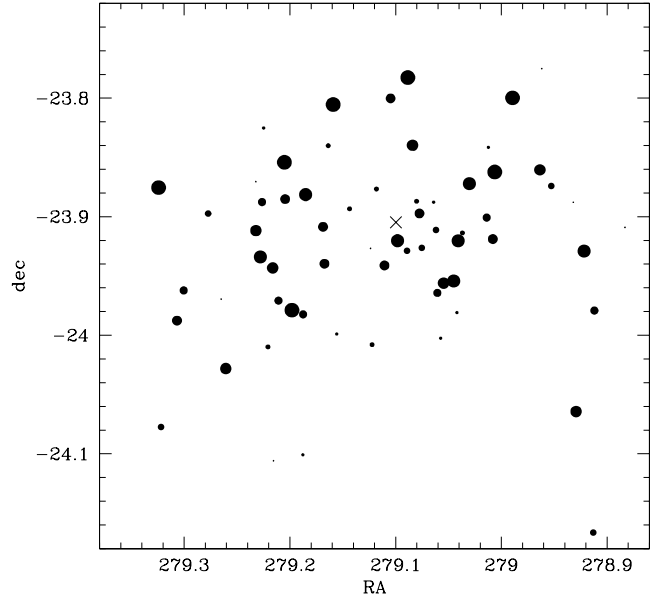}
\caption{Spatial distribution of the observed stars, where the size of each plotted point is proportional to the star's reddening. The cross indicates the cluster center.}
\label{redmap}
\end{center}
\end{figure}

The map of reddening in the cluster area is shown in Fig.~\ref{redmap}. The pattern is patchy, and the data do not reveal a clear gradient. In fact, the mean reddening calculated in the N/S and E/W halves and in the four quadrants do not differ more than 1$\sigma$ ($\approx$0.015~mag). If a spatial gradient of the reddening is present, it must be small, and probably blurred by small-scale variation. This is confirmed by the inspection of the detailed map of \citet{AlonsoGarcia12}, where they detect a clear pattern of reddening in the cluster area, but with limited variations around the mean value ($\approx\pm0.02$~mag). 

\subsection{Color-temperature relation}
\label{ss_colteff}

\begin{figure}
\begin{center}
\includegraphics[width=6.2cm,angle=270]{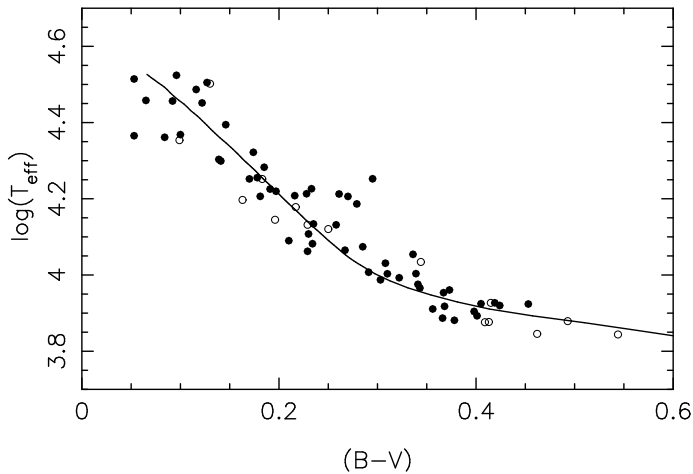}
\caption{Temperature-color relation for our target stars. Full and empty symbols are as in Fig.~\ref{f_cmd2}. The curve shows the best fit of the data, obtained shifting the theoretical relation of \citet{Pietrinferni06} and \citet{Cassisi09} by $E(B-V)$=0.34~mag.}
\label{colteff}
\end{center}
\end{figure}

The $T_\mathrm{eff}$--$(B-V)$ relation resulting from our measurements is shown in Fig.~\ref{colteff}. A quadratic fit of the data points returns the analytical expression
\begin{equation}
\log{(T_\mathrm{eff})}= 4.6594-2.6405\cdot(B-V)+1.9901\cdot(B-V)^2,
\label{eq_TeffBV}
\end{equation}
with a residual rms in $\log{(T_\mathrm{eff})}$ of 0.061~dex. A fit with the theoretical ZAHB color-temperature relation of \citet{Pietrinferni06} and \citet{Cassisi09} with Z=0.001 and Y=0.246 for the $\alpha$-enhanced mixture gives an independent estimate of the cluster reddening. We thus obtained $E(B-V)=0.34\pm0.02$~mag, consistent with what we found in Sect.~\ref{ss_reddening}. The theoretical curve, shifted horizontally by 0.34~magnitudes, reproduces the observed trend very well. \citetalias{moni2012} detected an anomalous behavior of $\omega$\,Cen HB stars in the $T_\mathrm{eff}$--$(U-V)$ plane, but not in the $T_\mathrm{eff}$--$(B-V)$ relation. Unfortunately, the photometric data of \citet{monaco2004} do not include the $U$ band, so we cannot check if this peculiarity is observed in M22 also.


\section{Conclusions}
\label{s_concl}

Our results on the gravities and the masses of HB stars in M22 with $T_\mathrm{eff}$=7\,000--25\,000~K match the previous measurements in NGC\,6752, M80, and NGC\,5986, which agree reasonably well with theoretical expectations. The anomalously low values found in $\omega$\,Cen thus remain a peculiarity of this cluster. Hence, while recent chemical studies highlighted that the stellar content of M22 presents many similarities with $\omega$\,Cen \citep{marino_milone09,marino2012}, its HB stars behave like their counterparts in the other ``normal" GCs. We detect a mild mass underestimate, intermediate between the results in $\omega$\,Cen and in the other clusters, for our targets hotter than 14\,000~K. This result is worth further investigation, but our data do not suggest a {\bf close} analogy with $\omega$\,Cen stars.

Five of our targets have a temperature compatible with objects hotter than the canonical end of the HB, but they all show low gravity and helium surface abundance, typical of post-HB stars. We therefore fail to detect any blue hook candidate, which to date have not been found in this cluster. The point is worth further investigation because, although our sample is too small and these very hot objects are hard to be identified in a highly reddened cluster such as M22, blue hook stars have been detected in most of the massive Galactic GCs \citep{Whitney98,Piotto99,DCruz00,Brown01,Rosenberg04,Momany04,Busso07,Ripepi07}.

We estimated the mean cluster reddening with two independent methods, finding $E(B-V)=0.36\pm0.02$ and $0.34\pm0.02$~mag. We can therefore fix our best estimate as $E(B-V)=0.35\pm0.02$~mag. The semi-amplitude of the maximum reddening variation is $\Delta(E(B-V))=\pm0.06$~mag, with a rms intrinsic dispersion of $\sigma(E(B-V))=0.03$~mag. These spectroscopic values are in excellent agreement with previous results available in the literature, obtained from the photometric properties of cluster red giants. Within the errors of our measurements, we do not detect a clear pattern or spatial gradient of the reddening in the cluster area.


\begin{acknowledgements}
DG and MC acknowledges support by the BASAL Center for Astrophysics and Associated Technologies (CATA) PFB-06/2007. Support for MC is also provided by the Chilean Ministry for the Economy, Development, and Tourism's Programa Inicativa Cient\'ifica Milenio through grant P07-021-F, awarded to The Milky Way Millennium Nucleus; by Proyecto FONDECYT Regular \#1110326; and by Proyecto Anillo de Investigaci\'on en Ciencia y Tecnolog\'ia PIA CONICYT-ACT 86. CMB thanks Dr. Maximiliano Moyano for his technical support. SV gratefully acknowledges the support provided by FONDECYT N. 1130721. We thank the anonymous referee for his/her detailed and constructive comments.
\end{acknowledgements}

\bibliographystyle{aa}
\bibliography{m22.bib}

\onllongtab{1}{
\begin{longtable}{lccccccccccccc}\\
\caption{Derived parameters of the target stars.}
\label{Tabla1}\\
 
\hline
ID & $V$ & $(B-V)$ & $T_\mathrm{eff}$ & $log(g)$ & & $log(\frac{N(He)}{N(H)})$ & & $E(B-V)$ & M & & $RV_H$ & & \\
& & & K & dex & & dex & & & $M_{\sun}$ & & km~s$^{-1}$ & & \\

\hline
\endfirsthead
 
\caption{continued.}\\
 \hline
ID & $V$ & $(B-V)$ & $T_\mathrm{eff}$ & $log(g)$ & & $log(\frac{N(He)}{N(H)})$ & & $E(B-V)$ & M & & $RV_H$ & & \\
\hline 
\endhead
\hline

2-91   & 14.339 & 0.368 & \ \ 9\,000 $\pm$\ \ 150    & 3.09 $\pm$ 0.12 & &       $-$        & & 0.381 & 0.37 $\pm$ 0.12 & & $-136$ & & \\
3-111  & 14.402 & 0.419 & \ \ 8\,300 $\pm$\ \ 100    & 3.18 $\pm$ 0.15 & &       $-$        & & 0.370 & 0.54 $\pm$ 0.29 & & $-142$ & & \\
2-936  & 14.893 & 0.303 & 10\,740 $\pm$\ \ 120       & 3.60 $\pm$ 0.09 & &       $-$        & & 0.367 & 0.48 $\pm$ 0.19 & & $-137$ & & \\
2-4213 & 17.088 & 0.083 & 23\,000 $\pm$3300          & 5.27 $\pm$ 0.09 & & $-3.10 \pm$ 0.29 & & 0.290 & 0.75 $\pm$ 0.40 & & $-139$ & & \\
2-2804 & 16.522 & 0.176 & 18\,000 $\pm$\ \ 900       & 4.65 $\pm$ 0.05 & & $-2.02 \pm$ 0.17 & & 0.349 & 0.50 $\pm$ 0.17 & &  $-96$ & \tiny{$\bigcirc$} & \\
1-217  & 15.215 & 0.209 & 12\,300 $\pm$\ \ 300       & 3.99 $\pm$ 0.05 & & $-2.06 \pm$ 0.28 & & 0.315 & 0.70 $\pm$ 0.29 & & $-126$ & & \\
2-2160 & 16.148 & 0.057 & 32\,700 $\pm$2400          & 5.19 $\pm$ 0.06 & & $-2.73 \pm$ 0.31 & & 0.326 & 0.86 $\pm$ 0.34 & & $-159$ & & \\
2-6658 & 17.544 & 0.043 & 23\,200 $\pm$2700          & 5,24 $\pm$ 0.09 & & $-1.77 \pm$ 0.09 & & 0.251 & 0.45 $\pm$ 0.23 & & $-145$ & & \\
3-73   & 14.055 & 0.531 & \ \ 6\,980 $\pm$\ \ \ \ 50 & 2.92 $\pm$ 0.13 & &       $-$        & & 0.279 & 4.16 $\pm$ 1.75 & $\bullet$ & $-132$ & & F \\
6-654  & 16.475 & 0.147 & 19\,900 $\pm$\ \ 500       & 4.76 $\pm$ 0.09 & & $-3.16 \pm$ 0.15 & & 0.335 & 0.55 $\pm$ 0.23 & &  $-91$ & \tiny{$\bigcirc$} & \\
1-158  & 14.881 & 0.275 & 11\,600 $\pm$\ \ 700       & 3.93 $\pm$ 0.02 & & $-1.69 \pm$ 0.28 & & 0.352 & 0.89 $\pm$ 0.26 & & $-123$ & & \\
2-2127 & 16.107 & 0.226 & 16\,300 $\pm$\ \ 500       & 4.46 $\pm$ 0.09 & & $-2.39 \pm$ 0.24 & & 0.379 & 0.56 $\pm$ 0.23 & & $-116$ & & \\
8-56   & 14.286 & 0.360 & \ \ 8\,100 $\pm$\ \ 100    & 3.02 $\pm$ 0.12 & &       $-$        & & 0.292 & 0.46 $\pm$ 0.21 & & $-140$ & & \\
2-495  & 14.273 & 0.355 & \ \ 8\,280 $\pm$\ \ 170    & 3.27 $\pm$ 0.12 & &       $-$        & & 0.301 & 0.77 $\pm$ 0.31 & & $-138$ & & \\
1-294  & 15.474 & 0.241 & 13\,500 $\pm$\ \ 600       & 4.13 $\pm$ 0.02 & & $-0.98 \pm$ 0.17 & \tiny{$\bigcirc$} & 0.365 & 0.62 $\pm$ 0.21 & &     0  & $\bullet$ & F \\
1-404  & 15.779 & 0.238 & 13\,600 $\pm$\ \ 900       & 4.04 $\pm$ 0.08 & & $-2.94 \pm$ 0.87 & & 0.367 & 0.39 $\pm$ 0.17 & & $-106$ & \tiny{$\bigcirc$} & \\
2-587  & 14.392 & 0.424 & \ \ 8\,450 $\pm$\ \ \ \ 40 & 3.01 $\pm$ 0.12 & &       $-$        & & 0.399 & 0.34 $\pm$ 0.16 & & $-136$ & & \\
2-5588 & 17.355 & 0.097 & 23\,400 $\pm$3400          & 5.01 $\pm$ 0.14 & & $-2.44 \pm$ 0.22 & & 0.310 & 0.31 $\pm$ 0.20 & & $-102$ & \tiny{$\bigcirc$} & \\
2-897  & 14.831 & 0.310 & 10\,080 $\pm$\ \ 130       & 3.43 $\pm$ 0.11 & &       $-$        & & 0.358 & 0.39 $\pm$ 0.17 & & $-135$ & & \\
2-8175 & 17.832 & 0.292 & 17\,900 $\pm$\ \ 400       & 4.34 $\pm$ 0.05 & & $-3.47 \pm$ 0.66 & & 0.468 & 0.07 $\pm$ 0.02 & \tiny{$\bigcirc$} & $-141$ & & \\
3-114  & 14.411 & 0.363 & \ \ 9\,140 $\pm$\ \ \ \ 50 & 3.31 $\pm$ 0.08 & &       $-$        & & 0.383 & 0.56 $\pm$ 0.21 & & $-155$ & & \\
3-225  & 15.038 & 0.481 & \ \ 7\,570 $\pm$\ \ 170    & 4.08 $\pm$ 0.11 & \tiny{$\bigcirc$} & $-$  & & 0.240 & 3.63 $\pm$ 1.50 & $\bullet$ & 11 & $\bullet$ & F \\
2-4766 & 17.175 & 0.191 & 16\,600 $\pm$\ \ 400       & 4.48 $\pm$ 0.09 & & $-1.82 \pm$ 0.24 & & 0.345 & 0.21 $\pm$ 0.09 & & $-120$ & & \\
3-71   & 14.053 & 0.382 & \ \ 7\,800 $\pm$\ \ 120    & 3.05 $\pm$ 0.08 & &       $-$        & & 0.269 & 0.71 $\pm$ 0.28 & & $-136$ & & \\
2-1111 & 15.143 & 0.229 & 11\,500 $\pm$\ \ 200       & 3.61 $\pm$ 0.02 & &       $-$        & & 0.315 & 0.35 $\pm$ 0.08 & & $-141$ & & \\
2-563  & 14.374 & 0.350 & \ \ 9\,240 $\pm$\ \ 110    & 3.37 $\pm$ 0.09 & &       $-$        & & 0.372 & 0.65 $\pm$ 0.26 & & $-133$ & & \\
2-6830 & 17.565 & 0.072 & 28\,700 $\pm$\ \ 800       & 4.34 $\pm$ 0.02 & & $-3.78 \pm$ 0.06 & & 0.330 & 0.04 $\pm$ 0.01 & \tiny{$\bigcirc$} & $-151$ & & \\
2-6777 & 17.537 & 0.085 & 28\,600 $\pm$2300          & 5.44 $\pm$ 0.12 & & $-1.11 \pm$ 0.67 & & 0.325 & 0.51 $\pm$ 0.27 & & $-116$ & & \\
1-2131 & 17.726 & 0.126 & 28\,300 $\pm$3000          & 5.26 $\pm$ 0.12 & & $-3.03 \pm$ 0.22 & & 0.351 & 0.29 $\pm$ 0.16 & & $-174$ & \tiny{$\bigcirc$} & \\
1-121  & 14.592 & 0.311 & \ \ 9\,720 $\pm$\ \ 110    & 3.50 $\pm$ 0.05 & &       $-$        & & 0.345 & 0.61 $\pm$ 0.18 & & $-138$ & & \\
2-4837 & 17.207 & 0.168 & 17\,900 $\pm$\ \ 500       & 4.73 $\pm$ 0.05 & & $-2.23 \pm$ 0.21 & & 0.339 & 0.32 $\pm$ 0.11 & & $-151$ & & \\
3-120  & 14.448 & 0.331 & 10\,800 $\pm$\ \ 200       & 3.75 $\pm$ 0.12 & &       $-$        & & 0.386 & 1.01 $\pm$ 0.41 & $\bullet$ & $-132$ & & F \\
3-270  & 15.292 & 0.221 & 12\,800 $\pm$\ \ 500       & 3.81 $\pm$ 0.02 & & $-2.44 \pm$ 0.73 & & 0.346 & 0.42 $\pm$ 0.11 & & $-127$ & & \\
8-82   & 14.599 & 0.413 & \ \ 7\,520 $\pm$\ \ 140    & 3.35 $\pm$ 0.15 & \tiny{$\bigcirc$} & $-$  & & 0.223 & 1.06 $\pm$ 0.38 & $\bullet$ & $-40$ & $\bullet$ & F \\
1-73   & 14.129 & 0.424 & \ \ 7\,530 $\pm$\ \ 120    & 3.38 $\pm$ 0.14 & \tiny{$\bigcirc$} & $-$  & & 0.233 & 1.69 $\pm$ 0.65 & $\bullet$ & $-130$ & & F \\
1-378  & 15.700 & 0.194 & 16\,800 $\pm$1400          & 4.37 $\pm$ 0.02 & & $-1.96 \pm$ 0.13 & & 0.357 & 0.62 $\pm$ 0.28 & & $-155$ & & \\
2-7157 & 17.612 & 0.141 & 24\,800 $\pm$3100          & 4.99 $\pm$ 0.09 & & $-2.24 \pm$ 0.24 & & 0.367 & 0.21 $\pm$ 0.10 & & $-146$ & & \\
2-6400 & 17.496 & 0.117 & 30\,700 $\pm$3900          & 5.53 $\pm$ 0.08 & & $-2.98 \pm$ 0.20 & & 0.367 & 0.58 $\pm$ 0.28 & & $-115$ & & \\
2-5785 & 17.495 & 0.129 & 32\,000 $\pm$\ \ 800       & 4.89 $\pm$ 0.06 & & $-3.27 \pm$ 0.69 & & 0.404 & 0.13 $\pm$ 0.04 & & $-122$ & & \\
2-2752 & 16.500 & 0.177 & 16\,100 $\pm$\ \ 300       & 4.38 $\pm$ 0.08 & & $-1.72 \pm$ 0.11 & & 0.329 & 0.34 $\pm$ 0.13 & & $-134$ & & \\
1-101  & 14.428 & 0.371 & \ \ 9\,500 $\pm$\ \ 200    & 3.15 $\pm$ 0.15 & &       $-$        & & 0.403 & 0.33 $\pm$ 0.18 & & $-136$ & & \\
2-4891 & 17.230 & 0.121 & 31\,800 $\pm$5500          & 4.65 $\pm$ 0.06 & & $-2.06 \pm$ 0.30 & & 0.398 & 0.09 $\pm$ 0.05 & \tiny{$\bigcirc$} & $-102$ & \tiny{$\bigcirc$} & F \\
3.169  & 14.729 & 0.283 & 10\,170 $\pm$\ \ 110       & 3.55 $\pm$ 0.13 & &       $-$        & & 0.331 & 0.56 $\pm$ 0.21 & & $-137$ & & \\
2-2427 & 16.292 & 0.255 & 16\,300 $\pm$\ \ 900       & 4.70 $\pm$ 0.05 & & $-1.67 \pm$ 0.13 & & 0.404 & 0.82 $\pm$ 0.31 & & $-103$ & \tiny{$\bigcirc$} & \\
2-528  & 14.320 & 0.404 & \ \ 8\,410 $\pm$\ \ \ \ 60 & 3.05 $\pm$ 0.12 & &       $-$        & & 0.376 & 0.42 $\pm$ 0.16 & & $-139$ & & \\
1-89   & 14.331 & 0.390 & \ \ 7\,710 $\pm$\ \ 110    & 2.95 $\pm$ 0.05 & &       $-$        & & 0.272 & 0.44 $\pm$ 0.17 & & $-126$ & & \\
2-3461 & 16.846 & 0.225 & 16\,800 $\pm$1500          & 4.35 $\pm$ 0.05 & & $-1.95 \pm$ 0.16 & & 0.385 & 0.20 $\pm$ 0.09 & & $-154$ & & \\
2-3509 & 16.863 & 0.276 & 15\,370 $\pm$\ \ 170       & 4.31 $\pm$ 0.02 & &       $-$        & & 0.423 & 0.22 $\pm$ 0.05 & & $-132$ & & \\
2-9272 & 17.903 & 0.184 & 21\,000 $\pm$3400          & 5.11 $\pm$ 0.09 & & $-2.89 \pm$ 0.58 & & 0.366 & 0.28 $\pm$ 0.15 & & $-115$ & & \\
2-476  & 14.233 & 0.394 & \ \ 8\,030 $\pm$\ \ 140    & 3.16 $\pm$ 0.11 & &       $-$        & & 0.313 & 0.70 $\pm$ 0.28 & & $-135$ & & \\
2-5018 & 17.262 & 0.089 & 22\,600 $\pm$1400          & 5.18 $\pm$ 0.09 & & $-2.77 \pm$ 0.24 & & 0.294 & 0.55 $\pm$ 0.25 & & $-220$ & $\bullet$ & F \\
1-211  & 15.172 & 0.353 & 11\,300 $\pm$\ \ 300       & 3.72 $\pm$ 0.08 & &       $-$        & & 0.429 & 0.43 $\pm$ 0.18 & & $-132$ & & \\
3-113  & 14.409 & 0.347 & \ \ 7\,610 $\pm$\ \ 150    & 3.04 $\pm$ 0.11 & &       $-$        & & 0.173 & 0.56 $\pm$ 0.31 & & $-130$ & & \\
3-163  & 14.696 & 0.410 & \ \ 8\,450 $\pm$\ \ 180    & 3.74 $\pm$ 0.09 & \tiny{$\bigcirc$} & $-$  & & 0.345 & 1.46 $\pm$ 0.60 & $\bullet$ & $-20$ & $\bullet$ & F \\
1-117  & 14.554 & 0.334 & \ \ 9\,840 $\pm$\ \ 160    & 3.44 $\pm$ 0.05 & &       $-$        & & 0.372 & 0.53 $\pm$ 0.20 & & $-122$ & & \\
1-147  & 14.779 & 0.357 & 10\,090 $\pm$\ \ 110       & 3.41 $\pm$ 0.11 & &       $-$        & & 0.405 & 0.38 $\pm$ 0.16 & & $-142$ & & \\
6-926  & 16.895 & 0.243 & 12\,100 $\pm$\ \ 900       & 3.72 $\pm$ 0.12 & & $-1.06 \pm$ 0.38 & & 0.370 & 0.08 $\pm$ 0.03 & \tiny{$\bigcirc$} & $-149$ & & \\
2-8090 & 17.743 & 0.147 & 15\,700 $\pm$3600          & 4.28 $\pm$ 0.03 & & $-1.66 \pm$ 0.49 & & 0.297 & 0.08 $\pm$ 0.04 & \tiny{$\bigcirc$} &   152  & $\bullet$ & F \\
1-429  & 15.861 & 0.265 & 13\,500 $\pm$\ \ 300       & 4.03 $\pm$ 0.08 & &       $-$        & & 0.394 & 0.37 $\pm$ 0.14 & & $-145$ & & \\
3-949  & 16.663 & 0.209 & 16\,200 $\pm$\ \ 900       & 4.37 $\pm$ 0.09 & & $-1.33 \pm$ 0.19 & & 0.361 & 0.29 $\pm$ 0.12 & & $-146$ & & \\
1-961  & 16.780 & 0.221 & 15\,100 $\pm$1400          & 4.09 $\pm$ 0.02 & & $-1.28 \pm$ 0.26 & \tiny{$\bigcirc$} & 0.368 & 0.15 $\pm$ 0.04 & & $-100$ & \tiny{$\bigcirc$} & F \\
1-1175 & 17.035 & 0.198 & 17\,900 $\pm$4700          & 4.91 $\pm$ 0.02 & & $-1.80 \pm$ 0.39 & & 0.360 & 0.55 $\pm$ 0.36 & & $-193$ & $\bullet$ & F \\
2-3338 & 16.815 & 0.088 & 33\,400 $\pm$2500          & 5.32 $\pm$ 0.06 & & $-3.96 \pm$ 0.15 & & 0.360 & 0.62 $\pm$ 0.27 & & $-124$ & & \\
1-146  & 14.774 & 0.244 & 13\,200 $\pm$1600          & 3.97 $\pm$ 0.08 & & $-1.16 \pm$ 0.57 & \tiny{$\bigcirc$} & 0.371 & 0.90 $\pm$ 0.44 & &  $-87$ & \tiny{$\bigcirc$} & F \\
2-3357 & 16.815 & 0.189 & 19\,200 $\pm$2000          & 4.82 $\pm$ 0.05 & & $-1.98 \pm$ 0.14 & & 0.369 & 0.49 $\pm$ 0.23 & & $-168$ & \tiny{$\bigcirc$} & \\
8-122  & 14.986 & 0.209 & 14\,000 $\pm$\ \ 600       & 4.42 $\pm$ 0.06 & & $-3.57 \pm$ 0.40 & & 0.331 & 1.82 $\pm$ 0.78 & $\bullet$ & $-135$ & & F \\
3-118  & 14.424 & 0.433 & \ \ 8\,390 $\pm$\ \ \ \ 90 & 3.19 $\pm$ 0.02 & &       $-$        & & 0.400 & 0.53 $\pm$ 0.21 & & $-129$ & & \\
1-498  & 16.042 & 0.278 & 16\,100 $\pm$\ \ 500       & 4.53 $\pm$ 0.08 & & $-2.08 \pm$ 0.21 & & 0.428 & 0.72 $\pm$ 0.29 & & $-142$ & & \\
1-1189 & 17.049 & 0.154 & 20\,100 $\pm$5900          & 4.90 $\pm$ 0.08 & & $-1.75 \pm$ 0.30 & & 0.341 & 0.42 $\pm$ 0.28 & & $-106$ & \tiny{$\bigcirc$} & \\
1-200  & 15.105 & 0.303 & 11\,900 $\pm$\ \ 400       & 3.71 $\pm$ 0.09 & &       $-$        & & 0.403 & 0.43 $\pm$ 0.18 & & $-117$ & & \\
1-71   & 14.090 & 0.458 & \ \ 7\,000 $\pm$\ \ 200    & 3.79 $\pm$ 0.05 & \tiny{$\bigcirc$} & $-$  & & 0.213 & 2.51 $\pm$ 1.00 & $\bullet$ & $-131$ & & F \\

\hline	
\end{longtable}
}

\end{document}